# Power Minimization Techniques in Distributed Base Station Antenna Systems using Non-Orthogonal Multiple Access

Joumana Farah, Antoine Kilzi, Charbel Abdel Nour, Catherine Douillard

*Abstract*—This paper introduces new approaches for combining non-orthogonal multiple access (NOMA) with distributed base station (DBS) deployments. The purpose of the study is to unlock the true potentials of DBS systems in the NOMA context, since all previous works dealing with power minimization in NOMA are performed in the CBS (centralized base station) context. This work targets a minimization of the total transmit power in each cell, under user rate and power multiplexing constraints. Different techniques are designed for the joint allocation of subcarriers, antennas and power, with a particular care given to insuring a moderate complexity. Results show an important gain in the total transmit power obtained by the DBS-NOMA combination, with respect to both DBS-OMA (orthogonal multiple access) and CBS-NOMA deployment scenarios.

## I. Introduction

The concept of distributed base stations (DBS) [1-2] was introduced in the past few years in mobile communication systems to increase the cell coverage in a cost effective way, and to strengthen the network infrastructure, particularly in saturated areas. It consists of deploying the base station (BS) antennas in a distributed manner throughout the cell, instead of having multiple antennas installed on a single tower at the cell center. The remote units, called remote radio heads (RRH) or remote antenna units (RAU), are connected to a processing and control center (PCC) through coax cables or fiber optics. By reducing the average distance of each mobile user to its transmitting/receiving antenna, the overall transmission power, necessary to ensure a certain quality of reception, is reduced in comparison to the centralized configuration (centralized base stations or CBS). Therefore, from the ecological standpoint, DBS can greatly reduce local electromagnetic radiation and $CO_2$ emissions of transmission systems. Alternatively, for the same overall transmission power as in CBS, DBS offers a higher and more uniform capacity over each cell. Moreover, it provides a better framework for improving system robustness to fading, intra-cell and inter-cell interferences, shadowing, and path loss. It also allows the system to better adapt to the varying user distribution. Besides, the use of DBS will allow the deployment of small antennas in large scale and in discrete locations in urban areas, e.g. on building roofs, electric poles, traffic and street lights, where they can be almost invisible due to their small size. This will significantly simplify and reduce the cost of site installation, therefore lowering the capital expenditure (CAPEX) of mobile operators.

Efficient implementation is key in squeezing the achievable potentials out of DBS systems. For this purpose, the study in [3] explored the advantages of DBS and compared the achievable ergodic capacity for two different transmission scenarios: selection diversity and blanket transmission. In the first one, one of the RRHs is selected (based on a path loss minimization criterion) for transmitting a given signal, whereas in the second, all antennas in the cell participate in each transmission, thus creating a macroscopic multiple antenna system. The results of this study show that selection diversity achieves a better capacity in the DBS context, compared to blanket transmission. The same observations are made in [4]. In [5], RRH selection is also preconized as a mean to decrease the number of information streams that need to be assembled from or conveyed to the involved RRHs, as well as the signaling overhead.

Several works target the optimization of system energy efficiency (EE) in the DBS context. In [6], two antenna selection techniques are proposed, either based on user pathloss information or RRH energy consumption. Also, proportional fairness scheduling is considered for subband allocation with a utility function adapted to optimize the EE. In [7], subcarrier assignment and power allocation are done in two separate stages. In the first one, the number of subcarriers per RRH is determined, and subcarrier/RRH assignment is performed assuming initial equal power distribution. In the second stage, power allocation (PA) is performed by maximizing the EE under the constraints of the total transmit power per RRH, the targeted Bit Error Rate and proportional fairness among users.

Moreover, non-orthogonal multiple access (NOMA) is currently being considered as a potential access scheme for 5G mobile communications. Several forms of NOMA are under evaluation [8]: Power-domain NOMA, sparse code multiple access (SCMA), multi-user shared access (MUSA), pattern division multiple access (PDMA), bit division multiplexing (BDM), ..., to name a few. This work targets power-domain NOMA, which applies the multiplexing of several users allocated to the same subcarrier in the power domain, by taking advantage of the channel gain difference between users [9-13]. At the receiver side, user separation is done using successive interference cancellation (SIC). Applying power multiplexing on top of the orthogonal frequency division multiplexing (OFDM) layer has proven to significantly increase system throughput compared to orthogonal signaling, while also improving fairness and cell-edge user experience.

In the majority of the early studies conducted on scheduling or resource allocation for NOMA like [10, 11], a proportional fairness (PF) scheduler is used to strike a balance between the average throughput and user fairness. Also, equal inter-subcarrier PA is assumed, while the repartition of power among multiplexed users on a subcarrier is often performed using fractional transmit power allocation (FTPA) [10, 11]. Despite its multiple advantages, the PF scheduler is not applicable in the context of power minimization, since it targets a tradeoff between total throughput and fairness, constrained by a fixed total transmission power.

In a former work [12], we have introduced a set of solutions for the problem of minimizing the spectrum occupancy in NOMA, under total BS power and user rate constraints. Also, in [13], an efficient method is proposed to incorporate a waterfilling inter-subcarrier PA within the PF scheduler.

J. Farah and A. Kilzi are with the Department of Electricity and Electronics, Faculty of Engineering, Lebanese University, Roumieh, Lebanon (joumana.farah@ul.edu.lb; kilzi.antoine@gmail.com).

C. Abdel Nour and C. Douillard are with Institut Mines-Telecom, CNRS UMR 6285 Lab-STICC, 29238 Brest, France, (email: charbel.abdelnour@imt-atlantique.fr; catherine.douillard@imt-atlantique.fr).

However, because of the difference in problem structure, both studies cannot be directly generalized to the case of power minimization. A few recent works tackle the power minimization problem in the NOMA context. In [14], a "relax-then-adjust" procedure is used to provide a suboptimal solution to the NP-hard problem: first, the problem is relaxed from the constraints relative to power domain multiplexing. Then, the obtained solution is iteratively adjusted using a bisection search, leading to a relatively high complexity. In [15], optimal PA is first conducted assuming fixed subcarrier assignment. Then, a deletion-based algorithm iteratively removes users from subcarriers until the constraints of the maximum number of multiplexed users are satisfied, thus necessitating a large number of iterations to converge. In [16], the authors propose an optimal and a suboptimal solution for determining the user scheduling, the SIC order, and the PA, for the case of a maximum of two users per subcarrier. However, the power domain multiplexing constraints are not taken into consideration. Power minimization strategies are also proposed in [17] for multiple-input multiple-output NOMA (MIMO-NOMA), where PA and receive beamforming design are alternated in an iterative way. Constraints on the targeted SINR (signal to interference and noise ratio) are considered to guarantee successful SIC decoding. The subcarrier allocation problem is not included, i.e., all users have access to the whole spectrum. Results, provided for a moderate number of users (4 or 6), show an important gain of performance with respect to orthogonal multiple access (OMA).

In [18], we have introduced a set of techniques that allow the joint allocation of subcarriers and power, with the aim of minimizing the total power in NOMA CBS. Particularly, we showed that the most efficient method, from the power perspective, consists of applying user pairing at a subsequent stage to single-user assignment, i.e., after applying OMA signaling as the first stage, instead of jointly assigning collocated users to subcarriers.

The main objective of this work is to study the potential of applying NOMA in the DBS context. To the best of our knowledge, the problem of power minimization in the DBS context has not been addressed in the literature: only the problem of EE optimization was considered in this context [6,7]. In fact, power minimization in DBS systems is a study item worth being explored since the established techniques for both OMA and NOMA in CBS do not simply extend to the generalized case of DBS. As it will be seen in this article, RRH selection, added to subcarrier assignment, user pairing, and PA, render the problem much more complex than in the CBS case. For this purpose, we will start by redesigning our previous CBS solution in [18], by exploiting particular properties of the waterfilling procedure, so as to decrease its complexity, without incurring any performance loss. Then, we will propose several solutions for extending the study to the DBS case. Most interestingly, we will show that by appropriately combining the pairing and RRH selection steps, and using certain information-theory properties of NOMA, two collocated users on a subcarrier can both perform SIC. The exploitation of such subcarriers can allow a significant performance enhancement, and therefore a particular care will be given for their allocation.

The paper is organized as follows: in Section II, we start by a description of the system model, with a formulation of the resource allocation problem in the context of DBS-NOMA. Then, in Section III, we present several suboptimal solutions for the power minimization problem, for the case of a single powering RRH per subcarrier. In Section IV, we introduce a novel approach for allowing a mutual SIC implementation on certain subcarriers, and introduce several allocation techniques for exploiting such subcarriers. Section V provides a brief overview of the complexity of the proposed algorithms. Section VI presents a performance analysis of the different allocation strategies, while Section VII concludes the paper.

## II. DESCRIPTION OF THE DBS-NOMA SYSTEM AND FORMULATION OF THE POWER MINIMIZATION PROBLEM

This study is conducted on a downlink system consisting of a total of $R$ RRHs uniformly positioned over a cell, where $K$ mobile users are randomly deployed. RRHs and users are assumed to be equipped with a single antenna. Users transmit channel state information (CSI) to RRHs, and the PCC collects all CSI from RRHs. Alternatively, in a TDD (time division duplexing) scenario, the PCC can benefit from channel reciprocity to perform channel estimation by exploiting uplink transmissions. Then, it allocates subcarriers, powers and RRHs to users in such a way to guarantee a transmission rate of $R_{k,req}$ [bps] for each user $k$. The system bandwidth $B$ is equally divided into a total of $S$ subcarriers. On the $n^{th}$ subcarrier $(1 \leq n \leq S)$, a maximum of $m(n)$ users $\{k_1, k_2, …, k_{m(n)}\}$ are chosen from the set of $K$ users, to be collocated on $n$ (or paired on $n$ when $m(n) = 2$). Classical OMA signaling corresponds to the special case of $m(n) = 1$. The framework is schematized in figure 1. NOMA subcarriers can be served by the same RRH or by different RRHs. For instance, one can consider serving User 1 and User 2 on the same subcarrier by RRH 1, while User 2 and User 3 are served (paired) on another subcarrier by RRH 1 and RRH 2 respectively.

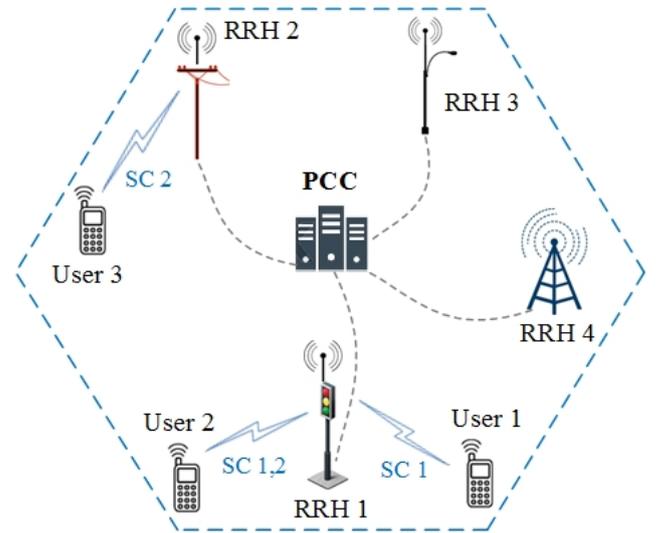

Fig. 1. Distributed Base Station system using NOMA (PCC = processing and control center, RRH = remote radio head, SC = subcarrier)

Let:

$P_{k_i,n,r}$ the power of the $i^{th}$ user on subcarrier $n$, transmitted by RRH $r$,

$h_{k_i,n,r}$ the channel coefficient between user $k_i$ and RRH $r$ over $n$,

$H$ the three-dimensional channel gain matrix with elements $h_{k,n,r}$, $1 \leq k \leq K$, $1 \leq n \leq S$, $1 \leq r \leq R$,

$N_0$ the power spectral density of additive white Gaussian noise, including randomized inter-cell interference, and assumed to be constant over all subcarriers.

A user $k_i$ on subcarrier $n$ can remove the inter-user interference from any other user $k_j$, collocated on $n$, whose channel gain verifies $h_{k_j,n} < h_{k_i,n}$ [9,10] and treats the received signals from other users as noise.

In the rest of the study, and without loss of generality, we will consider a maximum number of collocated users per subcarrier of 2, i.e., $m(n) = 1$ or 2. On the one hand, it has been shown [10] that the gain in performance obtained with the collocation of 3 users per subcarrier, compared to 2, is minor. On the other hand, limiting the number of multiplexed users per subcarrier reduces the SIC complexity in the receiver terminals. We will denote by first (resp. second) user the one having the higher (resp. lower) channel gain between the two users. Their theoretical throughputs $R_{k_i,n,r}$, $1 \leq i \leq 2$, on $n$ are given by the Shannon capacity limit as follows:

$$R_{k_1,n,r} = \frac{B}{S} \log_2 \left( 1 + \frac{P_{k_1,n,r} h^2_{k_1,n,r}}{N_0 B/S} \right), \tag{1}$$

$$R_{k_2,n,r} = \frac{B}{S} \log_2 \left( 1 + \frac{P_{k_2,n,r} h^2_{k_2,n,r}}{P_{k_1,n,r} h^2_{k_2,n,r} + N_0 B/S} \right), \tag{2}$$

**Proposition 2.1.** When the same RRH is used to power the signals of the two paired users on a subcarrier, only one of the two users is capable of performing SIC.

**Proof:** Let $R^{(k_1)}_{k_2,n,r}$ the necessary rate at user $k_1$ to decode the signal of user $k_2$:

$$R^{(k_1)}_{k_2,n,r} = \frac{B}{S} \log_2 \left( 1 + \frac{P_{k_2,n,r} h^2_{k_1,n,r}}{P_{k_1,n,r} h^2_{k_1,n,r} + N_0 B/S} \right).$$

Let $\sigma^2 = N_0 B/S$ the noise power on each subcarrier.

$k_1$ can perform SIC on $n$ if: $R^{(k_1)}_{k_2,n,r} \geq R_{k_2,n,r}$.

By writing: $R^{(k_1)}_{k_2,n,r} - R_{k_2,n,r} = \frac{B}{S} \log_2 \left( \frac{X}{Y} \right)$,

with:

$X = \left( \sigma^2 + P_{k_1,n,r} h^2_{k_1,n,r} + P_{k_2,n,r} h^2_{k_1,n,r} \right)\left( \sigma^2 + P_{k_1,n,r} h^2_{k_2,n,r} \right)$ and

$Y = \left( \sigma^2 + P_{k_1,n,r} h^2_{k_1,n,r} \right)\left( \sigma^2 + P_{k_1,n,r} h^2_{k_2,n,r} + P_{k_2,n,r} h^2_{k_2,n,r} \right)$.

After some calculations, we can write:

$X - Y = \sigma^2 P_{k_2,n,r} \left( h^2_{k_1,n,r} - h^2_{k_2,n,r} \right)$.

Therefore, $X - Y \geq 0$, since $h_{k_1,n,r} \geq h_{k_2,n,r}$.

Knowing that log(x) is a monotonically increasing function of x, this proves that $R^{(k_1)}_{k_2,n,r} - R_{k_2,n,r} \geq 0$

At the same time, $k_2$ cannot perform SIC since:

$$R^{(k_2)}_{k_1,n,r} - R_{k_1,n,r} = \frac{B}{S} \log_2 \left( 1 + \frac{P_{k_1,n,r} h^2_{k_2,n,r}}{P_{k_2,n,r} h^2_{k_2,n,r} + \sigma^2} \right) - R_{k_1,n,r}$$

$$= \frac{B}{S} \log_2 \left( \frac{Z}{T} \right)$$

with:

$Z - T = \sigma^2 P_{k_1,n,r} \left( h^2_{k_2,n,r} - h^2_{k_1,n,r} \right) - P_{k_1,n,r} P_{k_2,n,r} h^2_{k_1,n,r} h^2_{k_2,n,r} \leq 0.$ □

Proposition 2.1 will be of primary importance for the further development of our allocation techniques in the DBS-NOMA context.

Let $S_k$ be the set of subcarriers allocated to a user $k$, such that $k$ is either the first, second or sole user on any of its subcarriers, each of which being powered by a selected RRH. Let $T_k$ be the mapping set of RRHs corresponding to user $k$, such that the $i^{\text{th}}$ element of $T_k$ corresponds to the RRH selected for powering the $i^{\text{th}}$ subcarrier from set $S_k$. The corresponding optimization problem can be formulated as:

$$\{S_k, T_k, P_{k,n,r}\}^* = \underset{\{S_k, R_k, P_{k,n,r}\}}{\arg\min} \sum_{k=1}^{K} \sum_{\substack{n \in S_k \\ r=T_k(i), \text{ s.t. } S_k(i)=n}} P_{k,n,r},$$

subject to:

$$\sum_{n \in S_k} R_{k,n,r} = R_{k,\text{req}}, \ \forall \ 1 \leq k \leq K$$

$$P_{k,n,r} \geq 0, \ \forall \ 1 \leq k \leq K, n \in S_k, \ r \in T_k$$

$$P_{k_2,n,r} \geq P_{k_1,n,r}, \ \forall \ n \in S_k, 1 \leq k \leq K$$

The third constraint is the power multiplexing constraint proper to NOMA signaling.

The problem is a mixed combinatorial and non-convex one. Besides, compared to the case of NOMA CBS signaling, an additional dimension is added to the problem, which is the determination of the best RRH to power each allocated subcarrier to a user. In the sequel, we start by reviewing our previous solution in the NOMA CBS context. Then, we propose several enhancements to this solution, so as to pave the way for its adaptation to the DBS context.

### III. RESOURCE ALLOCATION TECHNIQUES FOR THE CASE OF A SINGLE POWERING RRH PER SUBCARRIER

*A. The previous power minimization technique for NOMA in CBS*

In [18], we showed that, in OMA signaling, the power allocation problem for a user $k$, over the set $S_k$ of its allocated subcarriers, can be formulated by a recursive low-complexity waterfilling technique. The latter provides the new waterline level for user $k$, after the assignment of a subcarrier $n$, as well as the power decrease $\Delta P_{k,n,r}$ incurred by this assignment. Let $N_k = \text{Card}(S_k)$ and $w_k(N_k)$ the corresponding waterline level. After adding a subcarrier $n_a$ to user $k$, the new waterline level, in terms of $w_k(N_k)$, is [18]:

$$w_k(N_k + 1) = \frac{\left( w_k(N_k) \right)^{N_k/N_k+1}}{\left( h^2_{k,n_a,r} / \sigma^2 \right)^{1/N_k+1}}. \tag{3}$$

Adding $n_a$ decreases the waterline only if its channel gain verifies [18]:

$$h^2_{k,n_a,r} > \frac{\sigma^2}{w_k(N_k)}. \tag{4}$$

The power decrease incurred by adding $n_a$ is expressed as:

$$\Delta P_{k,n,r} = (N_k + 1) w_k(N_k + 1) - N_k w_k(N_k) - \frac{\sigma^2}{h^2_{k,n_a,r}}. \tag{5}$$

Three main difficulties reside in applying NOMA in the power minimization context:

1. the achievable rate on each subcarrier being dependent on the user pairing order and on the inter-user interference term in the denominator of (2),
2. the necessity to meet $K$ independent user rate constraints,
3. the power domain multiplexing constraints that must be respected on each subcarrier to allow proper decoding at the receivers.

In [18], an efficient method was proposed for incorporating the waterfilling principle within NOMA signaling. It is summarized in Algorithm 1. Note that, since this algorithm was designed for the CBS case, in this part of the paper, $r$ designates the central (unique) BS antenna.

The initialization phase Worst-Best-H [18] is useful at the beginning of the algorithm to avoid depriving cell-edge users of their best subcarriers (essential in decreasing their power) in favor of cell-interior users. Afterwards, priority is based on the users' necessary total powers. The following notations are used: $S_k^{sole}$ is the set of subcarriers where user $k$ is the sole user (i.e., $m(n)=1, \forall n \in S_k^{sole}$), $R_k^{sole}$ the total rate of $k$ on subcarriers in $S_k^{sole}$, $S_k^{first}$ (resp. $S_k^{second}$) the set of subcarriers where $k$ is first (resp. second) user, collocated with a second (resp. first) one, $R_k^{first}$ and $R_k^{second}$ the total rates corresponding to $S_k^{first}$ and $S_k^{second}$, estimated using (1) and (2), $S_p$ the overall set of available subcarriers, $S_f$ the overall set of subcarriers assigned a first user without a second user, $P_{k,tot}$ the total amount of necessary power for user $k$ and $U_p$ the set of users whose power level can still be decreased (initially $U_p = \{1, 2, …, K\}$).

To estimate $\Delta P_{k_2,n,r}$ in phase 4 of Algorithm 1, the power needed on the subcarriers $S_{k_2}^{sole}$ is first found, constrained by $R_{k_2}^{sole}$. For this purpose, we calculate $R_{k_2}^{sole} = R_{k_2,req} - R_{k_2}^{first} - R_{k_2}^{second}$. Then, a gradual waterfilling is performed on the set $S_{k_2}^{sole}$, so as to reach $R_{k_2}^{sole}$ on this set. Following that, $\Delta P_{k_2,n,r}$ is estimated for candidate subcarrier $n$. Note that gradual dichotomy-based waterfilling is performed using the procedure described in [19]. Moreover, the allocation of subcarrier $n^*$ to $k_2$ may only decrease its power by a negligible amount. For this purpose, the power decrease is compared to a threshold $\rho$. The latter is chosen in such a way to strike a balance between power efficiency and spectral efficiency of the system, since unused subcarriers are released for use by other users or operators.

*Algorithm 1: NOMA-CBS*

**Phase 1:**
Attribute a subcarrier to each user using the Worst-Best-H priority

**Phase 2:** // Assigning first users to subcarrier using OMA signaling
$k^* = \arg\max_k P_{k,tot}$ // identify the user with the highest priority
**For** every $n \in S_p$ verifying (4)
 Calculate $w_{k^*}(N_{k^*}+1)$ using (3)
 Calculate $\Delta P_{k^*,n,r}$ using (5)
**End for**
$n^* = \arg\min_n \Delta P_{k^*,n,r}$
Attribute $n^*$ to $k^*$, unless $\Delta P_{k^*,n^*,r} > -\rho$ and update $P_{k^*,tot}$
**If** $\Delta P_{k^*,n^*,r} > -\rho$, remove $k^*$ from $U_p$
Repeat Phase 2 until no more subcarriers can be allocated

**Phase 3:**
Search for subcarriers with negative-powers (i.e., having $\sigma^2 / h_{k,n,r}^2 > w_k(N_k+1)$): free subcarriers one by one and update waterlines and powers

**Phase 4:** // Assigning second users to subcarriers using NOMA signaling, re-initialized by $U_p = \{1, 2, …, K\}$.
$k_2 = \arg\max_k P_{k,tot}$ // identify the user with the highest priority
**For** every $n \in S_f$ s.t. $h_{k_2,n,r} < h_{k_1,n,r}$ // $k_1$ is the first user on $n$
 Calculate $P_{k_2,n,r}$ using FTPA: $P_{k_2,n,r} = P_{k_1,n,r} h_{k_2,n,r}^{-2\alpha} / h_{k_1,n,r}^{-2\alpha}$, (6)
 Calculate $R_{k_2}^{first}$, $R_{k_2}^{second}$ and $R_{k_2}^{sole}$
 Perform gradual waterfilling on $S_{k_2}^{sole}$ constrained by $R_{k_2}^{sole}$
 // Calculate the new total power of $k_2$
 $P_{k_2,tot}^{(2)} = P_{k_2,n,r} + \sum_{n \in S_{k_2}^{sole}} P_{k_2,n,r} + \sum_{n \in S_{k_2}^{first}} P_{k_2,n,r} + \sum_{n \in S_{k_2}^{second}} P_{k_2,n,r}$
 $\Delta P_{k_2,n,r} = P_{k_2,tot}^{(2)} - P_{k_2,tot}^{(1)}$ // $P_{k_2,tot}^{(1)}$ is the previous total power of $k_2$
**End for**
$n^* = \arg\min_n \Delta P_{k_2,n,r}$
**If** $\Delta P_{k_2,n^*,r} < -\rho$
 Assign $k_2$ as second user on $n^*$
 $P_{k_1,n^*,r}$ and $P_{k_2,n^*,r}$ are fixed in the following iterations
**Else** free zero-power subcarriers of $k_2$ one by one, update its waterline and power levels; then, remove $k_2$ from $U_p$
Repeat Phase 4 until no more subcarriers can be allocated second users

*B. Runtime enhancement of the NOMA CBS solution and adaptation to the DBS context*

First, we start by revisiting the waterfilling principles summarized in section III.A, in order to introduce several procedures for reducing the complexity of Algorithm 1, prior to its adaptation to the DBS context (Algorithm 2). In this section, we consider the case where the first and second users on a subcarrier are powered by the same RRH, thus the name of the method NOMA-DBS-SRRH.

**Proposition 3.1.** In the OMA phase of NOMA-CBS (Cf. phase 2 of Algorithm 1), the subcarrier $n_a$ that ensures the lowest power decrease to user $k$ corresponds to the one with the highest channel gain among the available subcarriers, provided that it verifies (4).
**Proof:** Using (3), we can rewrite the power variation as:

$$\Delta P_{k,n,r} = (N_k+1)\left(\frac{(w_k(N_k))^{N_k}}{h_{k,n_a,r}^2/\sigma^2}\right)^{1/N_k+1} - N_k w_k(N_k) - \frac{\sigma^2}{h_{k,n_a,r}^2}. \quad (7)$$

By taking the derivative of $\Delta P_{k,n,r}$ with respect to $h_{k,n_a,r}$, we get:

$$\frac{\partial(\Delta P_{k,n,r})}{\partial(h_{k,n_a,r})} = -2\frac{(\sigma^2)^{1/N_k+1}}{(h_{k,n_a,r})^{\frac{2}{N_k+1}+1}}(w_k(N_k))^{N_k/N_k+1} + \frac{2\sigma^2}{h_{k,n_a,r}^3}.$$

Therefore, we can verify that:

$$\frac{\partial(\Delta P_{k,n,r})}{\partial(h_{k,n_a,r})} \leq 0 \Leftrightarrow \frac{\sigma^2}{h_{k,n_a,r}^2} \leq \left(\frac{\sigma^2}{h_{k,n_a,r}^2}\right)^{1/N_k+1}(w_k(N_k))^{N_k/N_k+1}, \quad (8)$$

which directly leads to (4).
We deduce that $\Delta P_{k,n,r}$ is a monotonically decreasing function of $h_{k,n_a,r}$, which concludes the proof of **Proposition 3.1**. □

**Proposition 3.1** also means that the subcarriers sequentially assigned by Algorithm 1 to a user $k$ are in decreasing order of channel gain (i.e., the first subcarrier assigned to $k$ has the highest channel gain in $S_k$). These results will allow us to significantly reduce the complexity of Phase 2 in Algorithm 1.

Furthermore, in case $h_{k,n_a,r}^2 = \frac{\sigma^2}{w_k(N_k)}$, one can verify using (7) that $\Delta P_{k,n,r} = 0$. Hence, any subcarrier verifying (4) not only reduces the waterline, as stated in [18], but also guarantees a decrease of the necessary power, i.e., $\Delta P_{k,n,r} < 0$.

**Proposition 3.2.** After assigning a subcarrier $n_a$ verifying (4) to a user $k$, all subcarriers of $k$, including $n_a$, have a positive power level (i.e., $w_k(N_k+1) > \frac{\sigma^2}{h_{k,n_a,r}^2}$).

**Proof:** the power allocated to subcarrier $n_a$ is:

$$P_{k,n_a,r} = w_k(N_k+1) - \frac{\sigma^2}{h_{k,n_a,r}^2}.$$

It can be written in terms of $w_k(N_k)$ as:

$$P_{k,n_a,r} = \frac{(w_k(N_k))^{N_k/N_k+1} - \left(\frac{\sigma^2}{h_{k,n_a,r}^2}\right)^{N_k/N_k+1}}{\left(\frac{h_{k,n_a,r}^2}{\sigma^2}\right)^{1/N_k+1}}.$$

This allows us to verify that $P_{k,n_a,r} > 0$ as long as (4) is verified, i.e., the allocated power to the added subcarrier $n_a$ is ensured to be positive. Besides, the power allocated to the subcarriers $n \in S_k$ of user $k$, after each waterline update, is higher for subcarriers with higher channel gains. Therefore, the most recently added subcarrier $n_a$ is the one with the lowest channel gain and the lowest allocated power amongst the subcarriers in $S_k$, due to the consequence of **Proposition 3.1**. Since this power is positive, the powers of all subcarriers in $S_k$ are positive. This concludes the proof of **Proposition 3.2** and allows us to completely rule out phase 3 in Algorithm 1. □

Next, we turn our attention to the pairing phase, i.e., the assignment of second users to subcarriers using NOMA. From the runtime perspective, the most constraining step in this phase of Algorithm 1 is the dichotomy-based waterfilling calculation. The latter aims at compensating for the additional rate brought to user $k$ by the newly added subcarrier $n$ (as second user) by removing this additional rate from the one that should be achieved on the sole subcarriers of $k$ (i.e., subcarriers in set $S_k^{sole}$). Therefore, we propose to replace this dichotomy-based waterline estimation by an efficient iterative waterline update as follows:

Recall that the total rate $R_k^{sole}$ to be distributed on the set $S_k^{sole}$ of user $k$ can be expressed as:

$$R_k^{sole} = \sum_{n \in S_k^{sole}} \frac{B}{S} \log_2\left(1 + \frac{P_{k,n,r} h_{k,n,r}^2}{\sigma^2}\right),$$

where $P_{k,n,r} = w_k(N_k^{sole}) - \frac{\sigma^2}{h_{k,n,r}^2}$, with $N_k^{sole} = \text{Card}(S_k^{sole})$, and $w_k(N_k^{sole})$ is the waterline on the sole subcarriers of $k$.

Therefore, $R_k^{sole}$ can be rewritten as:

$$R_k^{sole} = \sum_{n \in S_k^{sole}} \frac{B}{S} \log_2\left(\frac{w_k(N_k^{sole}) h_{k,n,r}^2}{\sigma^2}\right), \text{ leading to:}$$

$$w_k(N_k^{sole}) = \left(2^{R_k^{sole}S/B} \prod_{n \in S_k^{sole}} \frac{\sigma^2}{h_{k,n,r}^2}\right)^{1/N_k^{sole}}.$$

If a user $k$ is assigned as a second user to a subcarrier $n_s$, he will gain a rate on $n_s$, calculated using (6) and (2). This rate corresponds to the rate decrease $\Delta R_{k,n_s,r} < 0$ that should be compensated for on the sole subcarriers of $k$, so as to ensure the global rate constraint $R_{k,req}$. Let $R_k^{sole'} = R_k^{sole} + \Delta R_{k,n_s,r}$ the new rate to be distributed on $S_k^{sole}$. The corresponding new waterline can be expressed as:

$$w'_k(N_k^{sole}) = \left(2^{R_k^{sole'}S/B} \prod_{n \in S_k^{sole}} \frac{\sigma^2}{h_{k,n,r}^2}\right)^{1/N_k^{sole}}.$$

Therefore, it can be shown that:

$$w'_k(N_k^{sole}) = 2^{\frac{\Delta R_{k,n_s,r} S}{N_k^{sole} B}} w_k(N_k^{sole}). \quad (9)$$

The rate variation $\Delta R_{k,n_s,r}$ lowers the water level and may cause some subcarriers to have negative powers. Such subcarriers must be removed from $S_k^{sole}$. In the case where the change in waterline does not provoke any subcarrier removal, the resulting change in the total power of user $k$ is:

$$\Delta P_{k,n_s,r} = N_k^{sole}\left(w'_k(N_k^{sole}) - w_k(N_k^{sole})\right) + P_{k,n_s,r}. \quad (10)$$

When the waterline decrease induces the removal of some subcarriers, those subcarriers are the ones with the weakest channel gains in $S_k^{sole}$. Let us sort the elements of $S_k^{sole}$ in decreasing order of magnitude, that is, $h_{k,1,r(1)} > h_{k,2,r(2)} > \ldots > h_{k,N_k^{sole},r(N_k^{sole})}$, where by $r(i)$ we denote the RRH powering the $i^{th}$ subcarrier of user $k$ (those subcarriers are assumed to be numbered from 1 to $N_k^{sole}$). If the

last subcarrier is first removed, the resulting waterline is:
$$w_k\left(N_k^{sole}-1\right) = \left(w_k(N_k^{sole})\right)^{N_k^{sole}/N_k^{sole}-1}\left(h_{k,N_k,r}^2/\sigma^2\right)^{1/N_k^{sole}-1}.$$
Since $P_{k,N_k^{sole},r} < 0$, i.e., $h_{k,N_k^{sole},r}^2 w_k(N_k^{sole}) < \sigma^2$, we get:
$$w_k\left(N_k^{sole}-1\right) < w_k\left(N_k^{sole}\right).$$
This means that removing a subcarrier with a negative power always decreases the waterline. Therefore, any other subcarrier $n$ such that $h_{k,n,r(n)} > h_{k,N_k^{sole},r(N_k^{sole})}$ with an initially negative power before the removal of the last subcarrier will get an even more negative power after this removal. This leads us to the conclusion that the negative-power subcarriers can all be removed at once, rather than one by one as usually done in waterfilling algorithms [20]. The corresponding power variation is:
$$\Delta P_{k,L} = \left(N_k^{sole}-L\right)w_k\left(N_k^{sole}-L\right) - N_k^{sole}w_k\left(N_k^{sole}\right) + \sum_{j=0}^{L-1}\frac{\sigma^2}{h_{k,N_k-j,r(N_k-j)}^2} + P_{k,n_s,r}, \quad (11)$$
where $L$ is the number of negative-power subcarriers and $w_k\left(N_k^{sole}-L\right)$ the water level after the removal of $L$ subcarriers:
$$w_k\left(N_k^{sole}-L\right) = \left(w_k(N_k^{sole})\right)^{\frac{N_k^{sole}}{N_k^{sole}-L}}\prod_{j=0}^{L-1}\left(h_{k,N_k-j,r}^2/\sigma^2\right)^{\frac{1}{N_k^{sole}-L}} \quad (12)$$

After this removal some new negative-power subcarriers may arise that should be removed as well. However, statistical estimations on Monte-Carlo simulations of our algorithms have shown that negative-power subcarriers are very rare (less than one allocation case out of a thousand).

*Algorithm 2: NOMA-DBS-SRRH*

**Phase 1:**
Worst-Best-H subcarrier and RRH allocation

**Phase 2:** // single-user assignment
$k^* = \arg\max_k P_{k,tot}$ // identify the user with the highest priority
$(n^*, r^*) = \underset{(n,r),\text{ s.t. } n \in S_p \text{ and } (4)}{\arg\max} h_{k^*,n,r}$ // using proposition 3.1
Calculate $w_{k^*}(N_{k^*}+1)$ using (3) and $\Delta P_{k^*,n^*,r^*}$ using (5)
**If** $\Delta P_{k^*,n^*,r^*} < -\rho$ attribute $n^*$ to $k^*$, and update $P_{k^*,tot}$
**Else** remove $k^*$ from $U_p$
Repeat Phase 2 until no more subcarriers can be allocated

**Phase 3:** // NOMA pairing
$k_2 = \arg\max_k P_{k,tot}$
**For** every $n \in S_f$ s.t. $h_{k_2,n,r} < h_{k_1,n,r}$
  Calculate $P_{k_2,n,r}$ using (6) // $r$ is the RRH powering user $k_1$ on $n$
  Calculate $R_{k_2}^{first}$, $R_{k_2}^{second}$ and $R_{k_2}^{sole}$
  Calculate $\Delta P_{k_2,n,r}$ using (9) and (10)
**End for**
$n^* = \arg\min_n \Delta P_{k_2,n,r}$

**If** $\Delta P_{k_2,n^*,r} < -\rho$
  Assign $k_2$ on $n^*$
  Fix $P_{k_1,n^*,r^*}$ and $P_{k_2,n^*,r^*}$ and update $P_{k_2,n,r}$, $n \in S_{k_2}^{sole}$
**Else** free zero-power subcarriers of $k_2$ using (11) and (12);
  then, remove $k_2$ from $U_p$
Repeat Phase 3 until no more subcarriers can be allocated second users

*C. Enhancement of the NOMA DBS solution through local power optimization*

The power decrease incurred by a candidate subcarrier in the third phase of NOMA-DBS-SRRH is greatly influenced by the amount of power $P_{k_2,n,r}$ allocated to user $k_2$ on $n$ using FTPA. Indeed, the addition of a new subcarrier translates in a rise of the user power on the one hand, and in a power decrease due to the subsequent waterline reduction on his sole subcarriers (through (9)), on the other. Therefore, we propose to optimize the value of $P_{k_2,n,r}$ in such a way that the consequent user power reduction is minimized:
$$\underset{P_{k_2,n,r}}{\text{Min}} \Delta P_{k_2,n,r}$$
subject to: $P_{k_2,n,r} \geq P_{k_1,n,r}$ on $n$

By injecting (9) into (10), and expressing $\Delta R_{k_2,n,r}$ using (2), the corresponding Lagrangian can be written as:
$$L\left(P_{k_2,n,r},\lambda\right) = N_{k_2}^{sole} w_{k_2}\left(N_{k_2}^{sole}\right)\left(\left(1 + \frac{P_{k_2,n,r}h_{k_2,n,r}^2}{P_{k_1,n,r}h_{k_2,n,r}^2 + \sigma^2}\right)^{-\frac{1}{N_{k_2}^{sole}}} - 1\right)$$
$$+ P_{k_2,n,r} + \lambda\left(P_{k_2,n,r} - P_{k_1,n,r}\right)$$
where $\lambda$ is the Lagrange multiplier.
The corresponding Karush-Kuhn-Tucker (KKT) conditions are:
$$\begin{cases}-\frac{w_{k_2}\left(N_{k_2}^{sole}\right)h_{k_2,n,r}^2}{P_{k_1,n,r}h_{k_2,n,r}^2+\sigma^2}\left(1+\frac{P_{k_2,n,r}h_{k_2,n,r}^2}{P_{k_1,n}h_{k_2,n,r}^2+\sigma^2}\right)^{-\frac{1}{N_{k_2}^{sole}}-1} + 1 + \lambda = 0 \\ \lambda\left(P_{k_2,n,r}-P_{k_1,n,r}\right) = 0\end{cases}$$
We can verify that the second derivative of the Lagrangian is always positive, thus the corresponding solution constitutes a unique minimum. For $\lambda = 0$, this optimum is:
$$P_{k_2,n,r}^* = \left(\left(\frac{w_{k_2}\left(N_{k_2}^{sole}\right)h_{k_2,n,r}^2}{P_{k_1,n,r}h_{k_2,n,r}^2+\sigma^2}\right)^{\frac{N_{k_2}^{sole}}{N_{k_2}^{sole}+1}} - 1\right)\frac{P_{k_1,n,r}h_{k_2,n,r}^2+\sigma^2}{h_{k_2,n,r}^2} \quad (13)$$
For $\lambda \neq 0$, we get $P_{k_2,n,r} = P_{k_1,n,r}$.

However, in the latter case, a certain gap must be set between the power levels of the two paired users, in such a way to guarantee successful SIC decoding at the first user level. Indeed, a SINR level should be guaranteed to allow efficient SIC, as shown in [17]. Therefore, we will take:
$$P_{k_2,n,r} = P_{k_1,n,r}(1+\mu), \quad (14)$$
with $\mu$ a positive safety power margin that depends on practical SIC implementation. In other terms, if the obtained $P_{k_2,n,r}^*$ verifies the power constraint inequality, it is retained as the optimal solution; otherwise, it is taken as in (14). This method, referred to as "NOMA-DBS-SRRH-LPO", operates similarly to Algorithm 2, except that (6) in Phase 3 is replaced by either (13) or (14).

*D. NOMA DBS solution with optimal power allocation*

In this method, we propose to jointly optimize the inter-subcarrier and intra-subcarrier PA by applying the Relax-then-adjust procedure in [14] based on successive variable substitution. This technique is applied in our work subsequently to NOMA-DBS-SRRH-LPO, as shown in Algorithm 3.

---

*Algorithm 3: NOMA-DBS-SRRH-OPA*

**Phase 1:**
Apply NOMA-DBS-SRRH-LPO to determine first and second user assignments to subcarriers, as well as a provisional power allocation.

**Phase 2:**
Apply optimal PA using the procedure in [14].

---

## IV. RESOURCE ALLOCATION TECHNIQUES IN DBS FOR THE CASE OF MUTUAL SIC

*A. Theoretical foundation*

In this section, we consider the case where the users $k_1$ and $k_2$, collocated on subcarrier $n$, are powered by two different RRHs, respectively $r_1$ and $r_2$.

**Proposition 4.1.** Users $k_1$ and $k_2$ can both perform SIC, if:
$$h_{k_1,n,r_2} \geq h_{k_2,n,r_2} \tag{15}$$
$$h_{k_2,n,r_1} \geq h_{k_1,n,r_1} \tag{16}$$

The corresponding power multiplexing constraint is:
$$\frac{h_{k_1,n,r_1}^2}{h_{k_1,n,r_2}^2} \leq \frac{P_{k_2,n,r_2}}{P_{k_1,n,r_1}} \leq \frac{h_{k_2,n,r_1}^2}{h_{k_2,n,r_2}^2}. \tag{17}$$

**Proof:**
User $k_1$ can perform SIC on $n$ if $R_{k_2,n,r_2}^{(k_1)} \geq R_{k_2,n,r_2}$ with the following power multiplexing condition:
$$P_{k_1,n,r_1} h_{k_1,n,r_1}^2 \leq P_{k_2,n,r_2} h_{k_1,n,r_2}^2. \tag{18}$$

Note that if $r_1 = r_2 = r$, (18) reverts to the previous case of a single RRH per subcarrier, i.e., $P_{k_1,n,r} \leq P_{k_2,n,r}$.

Similarly, $k_2$ can perform SIC on $n$ if $R_{k_1,n,r_1}^{(k_2)} \geq R_{k_1,n,r_1}$, with
$$P_{k_2,n,r_2} h_{k_2,n,r_2}^2 \leq P_{k_1,n,r_1} h_{k_2,n,r_1}^2 \tag{19}$$

We can then identify the conditions that guarantee a mutual SIC, that is, both users performing SIC. In such conditions, the reachable rates by users $k_1$ and $k_2$ become:

$$R_{k_1,n,r_1} = \frac{B}{S} \log_2 \left( 1 + \frac{P_{k_1,n,r_1} h_{k_1,n,r_1}^2}{\sigma^2} \right) \tag{20}$$

$$R_{k_2,n,r_2} = \frac{B}{S} \log_2 \left( 1 + \frac{P_{k_2,n,r_2} h_{k_2,n,r_2}^2}{\sigma^2} \right) \tag{21}$$

Following the same reasoning as in the proof of **Proposition 2.1**, we can write:

$$R_{k_2,n,r_2}^{(k_1)} - R_{k_2,n,r_2} = \frac{B}{S} \log_2 \left( 1 + \frac{P_{k_2,n,r_2} h_{k_1,n,r_2}^2}{P_{k_1,n,r_1} h_{k_1,n,r_1}^2 + \sigma^2} \right) - R_{k_2,n,r_2}$$

$$= \frac{B}{S} \log_2 \left( \frac{X}{Y} \right)$$

where
$$X - Y = \sigma^2 P_{k_2,n,r_2} \left( h_{k_1,n,r_2}^2 - h_{k_2,n,r_2}^2 \right) - P_{k_1,n,r_1} P_{k_2,n,r_2} h_{k_1,n,r_1}^2 h_{k_2,n,r_2}^2$$

Similarly:
$$R_{k_1,n,r_1}^{(k_2)} - R_{k_1,n,r_1} = \frac{B}{S} \log_2 \left( \frac{Z}{T} \right), \text{ with}$$

$$Z - T = \sigma^2 P_{k_1,n,r_1} \left( h_{k_2,n,r_1}^2 - h_{k_1,n,r_1}^2 \right) - P_{k_1,n,r_1} P_{k_2,n,r_2} h_{k_1,n,r_1}^2 h_{k_2,n,r_2}^2$$

In practical transmission situations, the second term in the expression of $X$ - $Y$ is much smaller than the first, and the same goes for $Z$ - $T$. Therefore, the second terms can be neglected, leading to the conditions (15) and (16) for mutual SIC. □

In the special case where $r_1 = r_2 = r$, only one of the two conditions (15) and (16) can be verified at a time, i.e., only one of the two users can perform SIC. Otherwise, when (15) and (16) are both verified, we also have:
$$\frac{h_{k_1,n,r_1}^2}{h_{k_1,n,r_2}^2} \leq \frac{h_{k_2,n,r_1}^2}{h_{k_2,n,r_2}^2}.$$

Therefore, in light of (18) and (19), one can conclude that when (15) and (16) are simultaneously verified, a PA scheme can be found to allow a mutual SIC by ensuring (17).

*B. Optimal solution for the unconstrained case*

In the case where the power multiplexing constraints (18) and (19) are discarded, one can verify, using (20) and (21) and simple Lagrangian optimization, that the PA in the pairing phase reverts to the user-specific waterfilling solution, similarly to the phase 2 in NOMA-DBS-SRRH or NOMA-DBS-SRRH-LPO. The only difference resides in that only candidate subcarriers verifying (15) and (16) are considered for pairing. This technique, used as a lower-bound benchmark on the total power, will be referred to as "NOMA-DBS-MutSIC-UC".

*C. Optimal formulation for the constrained case*

The fully constrained case can be cast as the solution of the following optimization problem:

$$\max_{\{P_{k,n,r}\}} \left( -\sum_{k=1}^{K} \sum_{n=1}^{S} \sum_{r=1}^{R} P_{k,n,r} \right), \text{ subject to:}$$

$$\sum_{n \in S_k} \log_2 \left( 1 + \frac{P_{k,n,r} h_{k,n,r}^2}{\sigma^2} \right) = R_{k,\text{req}}, \ 1 \leq k \leq K$$

$$\frac{P_{k_2,n,r_2}}{P_{k_1,n,r_1}} \leq \frac{h_{k_2,n,r_1}^2}{h_{k_2,n,r_2}^2}, \forall n \in S_{mSIC}$$

$$-\frac{P_{k_2,n,r_2}}{P_{k_1,n,r_1}} \leq -\frac{h_{k_1,n,r_1}^2}{h_{k_1,n,r_2}^2}, \forall n \in S_{mSIC}$$

$$P_{k,n,r} \geq 0, \forall k,n,r$$

$S_{mSIC}$ is the set of subcarriers where mutual SIC is performed. The corresponding Lagrangian with multipliers $\lambda_k$ and $\beta_{i,n}$ is:

$$L(P,\lambda,\beta_1,\beta_2) = -\sum_{k=1}^{K}\sum_{n=1}^{S}\sum_{r=1}^{R} P_{k,n,r} + \sum_{n \in S_{mSIC}} \beta_{1,n}\left(\frac{h_{k_2,n,r_1}^2}{h_{k_2,n,r_2}^2} - \frac{P_{k_2,n,r_2}}{P_{k_1,n,r_1}}\right)$$

$$+ \sum_{n \in S_{mSIC}} \beta_{2,n}\left(\frac{P_{k_2,n,r_2}}{P_{k_1,n,r_1}} - \frac{h_{k_1,n,r_1}^2}{h_{k_1,n,r_2}^2}\right)$$

$$+ \sum_{k=1}^{K} \lambda_k \left(R_{k,req} - \sum_{n \in S_k} \log_2\left(1 + \frac{P_{k,n,r} h_{k,n,r}^2}{\sigma^2}\right)\right)$$

Writing the KKT conditions (not presented here for the sake of concision) leads to a system of $N_e$ non-linear equations with $N_e$ variables, where $N_e = 3\text{Card}(S_{mSIC})+K+S$ (taking into account the $S$-Card($S_{mSIC}$) power variables on non-paired subcarriers). Using the fact that $\beta_{1,n}$ and $\beta_{2,n}$ cannot be simultaneously non-zero on $n$, there are $3^{\text{Card}(S_{mSIC})}$ combinations to solve, with $2\text{Card}(S_{mSIC})+K+S$ variables. This leads to a prohibitive complexity; therefore, next we elaborate different resource allocation strategies in which we account for the power multiplexing constraints at every subcarrier assignment iteration.

### D. Suboptimal solution for the constrained case using direct power adjustment (DPA)

In this section, a simpler yet efficient power adjustment based technique is proposed, where the adjustment is carried out after every subcarrier allocation. Focusing on the pairing step, and following a similar reasoning to the one in III.C, we can write the power variation, due to the pairing of user $k_2$ on $n$ with mutual SIC, as:

$$\Delta P_{k_2,n,r_2} = N_{k_2}^{sole} w_{k_2}\left(N_{k_2}^{sole}\right)\left(\left(1 + \frac{P_{k_2,n,r_2} h_{k_2,n,r_2}^2}{\sigma^2}\right)^{-\frac{1}{N_{k_2}^{sole}}} - 1\right) + P_{k_2,n,r_2} \quad (22)$$

Using simple mathematical derivation, we can verify that:

$$\frac{\partial(\Delta P_{k_2,n,r_2})}{\partial(P_{k_2,n,r_2})} \geq 0 \Leftrightarrow P_{k_2,n,r_2} \geq \frac{w_{k_2}\left(N_{k_2}^{sole}\right)^{\frac{N_{k_2}^{sole}}{N_{k_2}^{sole}+1}}}{\left(h_{k_2,n,r_2}^2/\sigma^2\right)^{\frac{1}{N_{k_2}^{sole}+1}}} - \frac{\sigma^2}{h_{k_2,n,r_2}^2} = P_{k_2,n,r_2}^*$$

In fact, $P_{k_2,n,r_2}^*$ is the power corresponding to the waterfilling applied to the candidate subcarrier $n$ and the sole subcarriers of user $k_2$. Also, we can verify that the second derivative of $\Delta P_{k_2,n,r_2}$ with respect to $P_{k_2,n,r_2}$ is always positive. Therefore, we deduce that for any value of $P_{k_2,n,r_2}$ greater (resp. lower) than $P_{k_2,n,r_2}^*$, $\Delta P_{k_2,n,r_2}$ is strictly increasing (resp. decreasing) with $P_{k_2,n,r_2}$. Consequently, the further $P_{k_2,n,r_2}$ is apart from $P_{k_2,n,r_2}^*$, the greater is $\Delta P_{k_2,n,r_2}$. Hence, the best choice of $P_{k_2,n,r_2}^*$, when it does not verify (17), should be at the limits of the inequality (17). This leads us to the suboptimal Algorithm 4.

### Algorithm 4: NOMA-DBS-MutSIC-DPA

**Phase 1:**
Worst-Best-H subcarrier and RRH allocation followed by OMA single-user assignment

**Phase 2:** // NOMA pairing
$k_2 = \arg\max_{k} P_{k,tot}$

$S_c = \{(n,r_2) \text{ s.t. } (15),(16) \& (4) \text{ are verified}\}$

**For** every candidate couple $(n,r_2) \in S_c$

    Calculate $P_{k_2,n,r_2}$ and $\Delta P_{k_2,n,r_2}$ using (3) and (5)

    **If** $P_{k_2,n,r_2}$ verifies (17), set $P_{k_2,n,r_2}^* = P_{k_2,n,r_2}$

    **If** $\frac{P_{k_2,n,r_2}}{P_{k_1,n,r_1}} < \frac{h_{k_1,n,r_1}^2}{h_{k_1,n,r_2}^2}$, set $P_{k_2,n,r_2}^* = P_{k_1,n,r_1}\frac{h_{k_1,n,r_1}^2}{h_{k_1,n,r_2}^2}(1+\mu)$ and

    estimate $\Delta P_{k_2,n,r_2}$ using (9) and (10)

    **If** $\frac{P_{k_2,n,r_2}}{P_{k_1,n,r_1}} > \frac{h_{k_2,n,r_1}^2}{h_{k_2,n,r_2}^2}$, set $P_{k_2,n,r_2}^* = P_{k_1,n,r_1}\frac{h_{k_2,n,r_1}^2}{h_{k_2,n,r_2}^2}(1-\mu)$ and

    estimate $\Delta P_{k_2,n,r_2}$ using (9) and (10)

**End for**

$(n^*,r_2^*) = \arg\min_{(n,r_2)} \Delta P_{k_2,n,r_2}$

Continue similarly to NOMA-DBS-SRRH

### E. Suboptimal solutions for the constrained case using sequential optimization for power adjustment

In order to further optimize the NOMA-DBS-MutSIC-DPA technique, we propose to replace the adjustment and power estimation steps by a sequential power optimization. Instead of optimizing the choice of $P_{k_2,n,r_2}$ over the candidate couple $(n, r_2)$, we look for a wider optimization in which powers of both first and second users on the considered subcarrier are adjusted, in a way that their global power variation is optimal:

$$\{P_{k_1,n,r_1}, P_{k_2,n,r_2}\}^* = \arg\max_{P_{k_1,n,r_1}, P_{k_2,n,r_2}} \left(-\Delta P = -\Delta P_{k_1,n,r_1} - \Delta P_{k_2,n,r_2}\right)$$

subject to:

$$\frac{P_{k_2,n,r_2}}{P_{k_1,n,r_1}} \geq \frac{h_{k_1,n,r_1}^2}{h_{k_1,n,r_2}^2}, \quad \frac{P_{k_2,n,r_2}}{P_{k_1,n,r_1}} \leq \frac{h_{k_2,n,r_1}^2}{h_{k_2,n,r_2}^2}, \quad P_{k_1,n,r_1} \geq 0, P_{k_2,n,r_2} \geq 0$$

$\Delta P_{k_2,n,r_2}$ is expressed as in (22), while $\Delta P_{k_1,n,r_1}$ is given by:

$$\Delta P_{k_1,n,r_1} = \left(N_{k_1}^{sole}-1\right) W_{I,k_1}\left(2^{-\frac{\Delta R_{k_1,n,r}S}{\left(N_{k_1}^{sole}-1\right)B}} - 1\right) + P_{k_1,n,r_1} - P_{k_1,n,r_1}^I$$

where $P_{k_1,n,r_1}^I$ is the initial power allocated on $n$ to $k_1$ and $W_{I,k1}$ the initial waterline of $k_1$ (before pairing with user $k_2$ and power adjustment). Also, the rate variation of user $k_1$ on $n$ can be written as:

$$\Delta R_{k_1,n,r} = \frac{B}{S}\log_2\left(\frac{\sigma^2 + P_{k_1,n,r_1} h_{k_1,n,r_1}^2}{\sigma^2 + P_{k_1,n,r_1}^I h_{k_1,n,r_1}^2}\right).$$

The Lagrangian of this problem is:

$$L(P_{k_1,n,r_1}, P_{k_2,n,r_2}, \lambda_1, \lambda_2) = -\Delta P_{k_1,n,r_1} - \Delta P_{k_2,n,r_2}$$

$$-\lambda_1\left(P_{k_1,n,r_1}\frac{h_{k_1,n,r_1}^2}{h_{k_1,n,r_2}^2} - P_{k_2,n,r_2}\right) - \lambda_2\left(P_{k_2,n,r_2} - P_{k_1,n,r_1}\frac{h_{k_2,n,r_1}^2}{h_{k_2,n,r_2}^2}\right)$$

The solution of this problem must verify the following conditions:

$$\nabla L(P_{k_1,n,r_1}, P_{k_2,n,r_2}, \lambda_1, \lambda_2) = 0$$
$$\lambda_1 (P_{k_1,n,r_1} h_{k_1,n,r_1}^2 / h_{k_1,n,r_2}^2 - P_{k_2,n,r_2}) = 0$$
$$\lambda_2 (P_{k_2,n,r_2} - P_{k_1,n,r_1} h_{k_2,n,r_1}^2 / h_{k_2,n,r_2}^2) = 0$$
$$\lambda_1, \lambda_2 \geq 0$$
$$P_{k_1,n,r_1}, P_{k_2,n,r_2} \geq 0$$

Four cases are identified:
1. $\lambda_1 = 0, \lambda_2 = 0$
2. $\lambda_1 \neq 0, \lambda_2 = 0 \rightarrow P_{k_2,n,r_2} = P_{k_1,n,r_1} h_{k_1,n,r_1}^2 / h_{k_1,n,r_2}^2$
3. $\lambda_1 = 0, \lambda_2 \neq 0 \rightarrow P_{k_2,n,r_2} = P_{k_1,n,r_1} h_{k_2,n,r_1}^2 / h_{k_2,n,r_2}^2$
4. $\lambda_1 \neq 0, \lambda_2 \neq 0$

Case 1 corresponds to the unconstrained waterfilling solution applied separately to the two concerned users. Case 4 is generally impossible, since the two boundaries of the inequality (17) would be equal. Considering Case 2, by replacing $P_{k_2,n,r_2}$ in terms of $P_{k_1,n,r_1}$ in the Lagrangian and by taking the derivative with respect to $P_{k_1,n,r_1}$, we can verify that $P_{k_1,n,r_1}^*$ is the solution of the following nonlinear equation:

$$\frac{W_{I,k_1} h_{k_1,n,r_1}^2}{\sigma^2 + P_{k_1,n,r_1}^I h_{k_1,n,r_1}^2} \left( \frac{\sigma^2 + P_{k_1,n,r_1} h_{k_1,n,r_1}^2}{\sigma^2 + P_{k_1,n,r_1}^I h_{k_1,n,r_1}^2} \right)^{-\frac{1}{N_{k_1}^{sole}-1}-1} - \frac{h_{k_1,n,r_1}^2}{h_{k_1,n,r_2}^2} - 1$$
$$+ W_{I,k_2} \frac{h_{k_1,n,r_1}^2 h_{k_2,n,r_2}^2}{h_{k_1,n,r_2}^2 \sigma^2} \left( 1 + \frac{P_{k_1,n,r_1} h_{k_2,n,r_2}^2 h_{k_1,n,r_1}^2}{\sigma^2} \right)^{-\frac{1}{N_{k_2}^{sole}-1}-1} = 0 \quad (23)$$

Note that in practice, we also take into consideration the safety power margin $\mu$ in the calculation of $P_{k_1,n,r_1}$. Similar calculations are performed for Case 3. The solution that yields the lowest $\Delta P$ is retained. Also, if none of the cases provides positive power solutions, the current candidate couple $(n, r_2)$ is discarded. This method of optimal power adjustment (OPAd) will be referred to as "NOMA-DBS-MutSIC-OPAd".

In order to decrease the complexity of "NOMA-DBS-MutSIC-OPAd", inherent to the resolution of nonlinear equations, we consider a semi-optimal variant of this technique, called "NOMA-DBS-MutSIC-SOPAd": at the stage where candidate couples $(n, r_2)$ are considered for potential assignment to user $k_2$, DPA is used for power adjustment, in order to determine the best candidate in a cost-effective way. Then, the preceding OPAd solution is applied to allocate power levels to users $k_1$ and $k_2$ on the retained candidate.

*F. Combination of the allocation of mutual and single SIC subcarriers in DBS*

The case of two different powering RRHs per subcarrier with only one user performing SIC is studied based on information theory developments similar to the ones performed in section IV.A. For instance, if (15) is verified and (16) is not, $k_1$ performs SIC on subcarrier $n$, while $k_2$ does not. The corresponding power multiplexing conditions become:

$$P_{k_1,n,r_1} h_{k_1,n,r_1}^2 \leq P_{k_2,n,r_2} h_{k_1,n,r_2}^2 \text{ and } P_{k_2,n,r_2} h_{k_2,n,r_2}^2 \geq P_{k_1,n,r_1} h_{k_2,n,r_1}^2.$$

In order to further exploit the space diversity inherent to DBS systems, we propose to first apply NOMA-DBS-MutSIC-SOPAd in order to identify and allocate subcarriers allowing mutual SIC. Then, in a subsequent phase, the remaining set of solely assigned subcarriers is further examined for potential allocation of a second user, using either the same or a different RRH from that of the first assigned user, but such that only the latter performs SIC. LPO is used for power allocation in this second phase. This method will be referred to as "NOMA-DBS-Mut&SingSIC".

V. COMPLEXITY ANALYSIS

In this section, we analyze the complexity of the different allocation techniques proposed in this study. The complexity of OMA-CBS, NOMA-CBS and OMA-DBS is studied by considering an implementation that includes the runtime enhancement procedures introduced in section III.B.

Starting with OMA-CBS, we consider that the channel matrix is reordered such as for each user the subcarriers are sorted by the decreasing order of channel gain. This step, that accelerates the subsequent subcarrier allocation stages, has a complexity of $O(KS\log(S))$. Following the Worst-Best-H phase, each iteration complexity is mainly dominated by the search of the most power consuming user with a cost $O(K)$. This leads to a total of $O(KS\log(S)+(S-K)K)$.

Each allocation step in the pairing phase of NOMA-CBS consists of the identification of the most power consuming user, followed by a search over the subcarrier space, and a power update over the set of the user's sole subcarriers, with an average number of $S/K$ subcarriers. Therefore, the total complexity of NOMA-CBS is $O(KS\log(S)+(S-K)K+S(K+S+S/K))$.

In OMA-DBS, we consider an initial sorting of each user subcarrier gains, separately for each RRH, with a cost of $O(KSR\log(S))$. Then, an allocation cycle will consist of user identification, followed by the search of the RRH providing the subcarrier with the highest channel gain. This corresponds to a complexity of $O(K+R)$. Therefore, the total complexity is: $O(KSR\log(S)+(S-K)(K+R))$. Consequently, the total complexity of NOMA-DBS-SRRH and NOMA-DBS-SRRH-LPO is $O(KSR\log(S)+(S-K)(K+R)+S(K+S+S/K))$.

The most constraining part in NOMA-DBS-SRRH-OPA is the resolution of a set of $N_{OPA}$ non-linear equations with $N_{OPA}$ unknowns, in the phase 2 of Algorithm 3. $N_{OPA} = \text{Card}(S_{mux})+K+S$, where $S_{mux}$ is the set of NOMA multiplexed subcarriers. Therefore, the complexity of this algorithm is $f_{comp}(N_{OPA})$, where $f_{comp}$ is a function that could either be exponential or polynomial in terms of $N_{OPA}$, depending on the resolution method.

Concerning NOMA-DBS-MutSIC-UC, by following the same reasoning as for OMA-DBS, and accounting for the search of an eventual collocated user for at most $S$ subcarriers, we get a total of $O(KSR\log(S)+(S-K)(K+R)+S(K+R))$.

As for NOMA-DBS-MutSIC-DPA, the total complexity is $O(KSR\log(S)+(S-K)(K+R)+S(K+S(R-1)+S/K))$, where the $S(R-1)$ term stems from the fact that the search over the subcarrier space in the pairing phase is conducted over all combinations of subcarriers and RRHs, except for the RRH of the first user on the candidate subcarrier.

In NOMA-DBS-MutSIC-OPAd, let C be the complexity of solving the nonlinear equation (23). The total complexity is therefore $O(KSR\log(S)+(S-K)(K+R)+S(K+S(R-1)C+S/K))$.

Given that NOMA-DBS-MutSIC-SOPAd solves (23) only once per allocation step, its complexity is O($KSR$log($S$)+($S$-$K$)($K$+$R$)+$S$($K$+$S$($R$-1)+$S$/$K$+$C$)). Consequently, the complexity of NOMA-DBS-Mut&SingSIC is O($KSR$log($S$)+($S$-$K$)($K$+$R$))+$S$($K$+$S$($R$-1)+$S$/$K$+$C$)+$SR$($K$+$S$+$S$/$K$). The additional term corresponds to the Single SIC phase which is similar to the pairing phase in NOMA-CBS except that the search space is enlarged by a factor $R$.

Table 1 summarizes the approximate complexity of the different techniques.

TABLE. 1. Approximate complexity of the different allocation techniques.

| Technique | Complexity |
|---|---|
| OMA-CBS | O($KS$log($S$)) |
| NOMA-CBS | O($S^2$+$KS$log($S$)) |
| OMA-DBS | O($KSR$log($S$)) |
| NOMA-DBS-SRRH | O($S^2$+$KSR$log($S$)) |
| NOMA-DBS-SRRH-LPO | O($S^2$+$KSR$log($S$)) |
| NOMA-DBS-SRRH-OPA | $f_{comp}(N_{OPA})$ |
| NOMA-DBS-MutSIC-UC | O($KSR$log($S$)) |
| NOMA-DBS-MutSIC-DPA | O($S^2R$) |
| NOMA-DBS-MutSIC-OPAd | O($S^2RC$) |
| NOMA-DBS-MutSIC-SOPAd | O($S^2R$+$SC$) |
| NOMA-DBS-Mut&SingSIC | O($S^2R$+$SC$) |

## VI. PRACTICAL RESULTS

The performance of the different allocation techniques are assessed through intensive simulations in the LTE/LTE-Advanced context [21]. The cell model is a hexagonal one with a radius $R_d$ of 500 m. For the DBS system, we consider a number $R$ of RRHs of 4 or 7. In each case, one antenna is located at the cell center, while the others are equally distanced and positioned on a circle of radius $2R_d/3$ centered at the cell center. The system bandwidth $B$ is 10 MHz. The transmission medium is a frequency-selective Rayleigh fading channel with an rms of 500 ns. We consider distance-dependent path loss with a decay factor of 3.76 and lognormal shadowing with an 8 dB variance. The noise power spectral density $N_0$ is $4.10^{-18}$ mW/Hz. In this study, we assume perfect knowledge of the user channel gains by the PCC. The $\alpha$ decay factor in FTPA is taken equal to 0.5, while the power threshold $\rho$ is 0.01 Watt as in [18]. The safety power margin $\mu$ is set to 0.01. OMA-CBS and OMA-DBS scenarios are also shown for comparison, where only phases 1 and 2 of Algorithm 2 are applied, using either $R$=1 (for OMA-CBS) or $R$≠1 (for OMA-DBS).

Figure 2 represents the total transmit power in the cell in terms of the requested rate, for the case of 15 users, 64 subcarriers and 4 RRHs. It shows that the DBS configuration greatly outperforms CBS: a large leap in power with a factor around 16 is achieved with both OMA and NOMA signaling. At a target rate of 12 Mbps, the required total power using OMA-DBS, NOMA-DBS-SRRH, NOMA-DBS-SRRH-LPO and NOMA-DBS-SRRH-OPA is respectively 39.79, 32.34, 30.04 and 29.41 W. This shows a clear advantage of NOMA over OMA in the DBS context. Besides, applying LPO allows a power reduction of 7.7% over FTPA, with a similar computational load, while the margin over optimal PA is of only 2% at 12 Mbps.

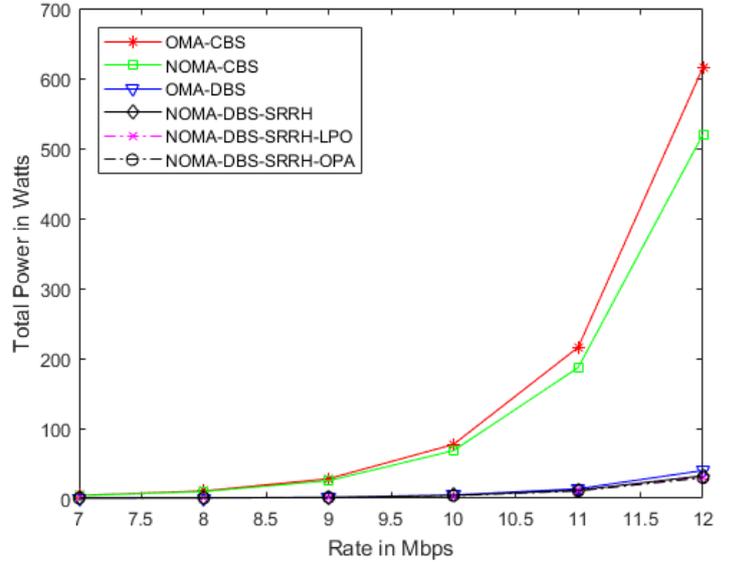

Fig. 2. Total power in terms of $R_{k,req}$, for $K$ = 15, $S$ = 64 and $R$ = 4.

We now turn our attention to the evaluation of mutual SIC and single SIC configurations. Figure 3 shows that all three constrained configurations based on pure mutual SIC (NOMA-DBS-MutSIC-DPA, NOMA-DBS-MutSIC-SOPAd and NOMA-DBS-MutSIC-OPAd) largely outperform NOMA-DBS-SRRH-LPO. Their gain towards the latter is respectively 10.91, 16.18 and 22.19 W, at a requested rate of 13 Mbps. The significant gain of optimal power adjustment towards its suboptimal counterpart comes at the cost of a significant complexity increase, as shown in Section V. The most power-efficient mutual SIC implementation is obviously NOMA-DBS-MutSIC-UC, since it is designed to solve a relaxed version of the power minimization problem, by dropping all power multiplexing constraints. Therefore, it essentially serves as a benchmark for assessing the other methods, because power multiplexing conditions are essential for allowing correct signal decoding at the receiver side.

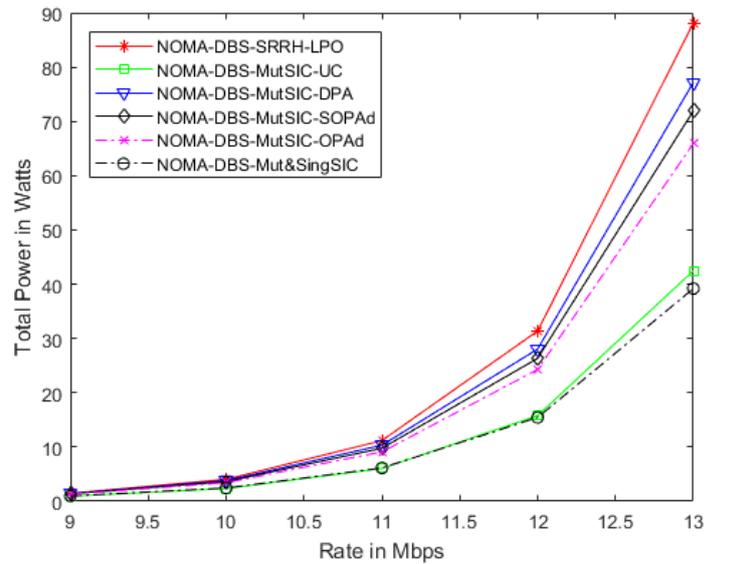

Fig. 3. Total power in terms of $R_{k,req}$, for $K$ = 15, $S$ = 64 and $R$ = 4.

The best global strategy remains the combination of mutual and single SIC subcarriers, since it allows a power reduction of 10.9 and 32.73 W, respectively at 12 and 13 Mbps, towards NOMA-DBS-MutSIC-SOPAd.

Figure 4 shows the influence of increasing the number of RRHs on the system performance. As expected, increasing the number of spread antennas greatly reduces the overall power, either with single SIC or combined mutual and single SIC configurations. A significant leap in power reduction is observed when $R$ is increased from 4 to 5, followed by a more moderate one when going from 5 to 7 antennas, and the same behavior is expected for larger values of $R$. However, considerations of practical order would suggest limiting the number of deployed RRHs in the cell to a certain extent, mainly because of the inherent overhead of CSI signaling exchange, not to mention geographical deployment constraints.

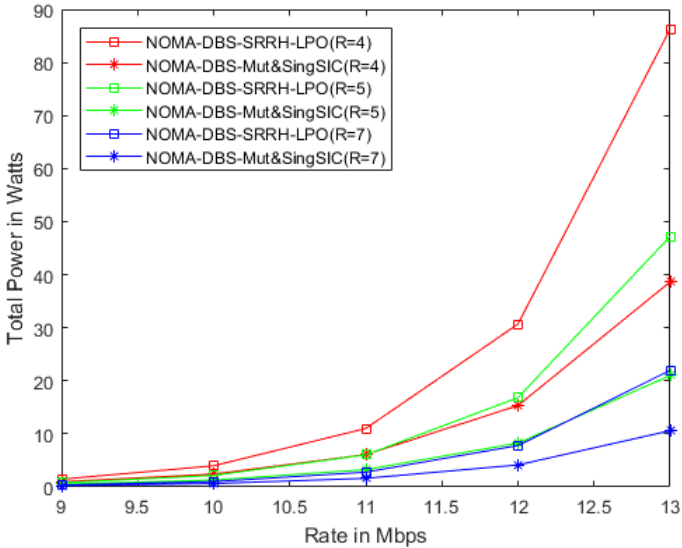

Fig. 4. Total power in terms of $R_{k,req}$, for $K = 15$, $S = 64$ and $R = 4$, 5 or 7.

In figure 5, we show the performance for a varying number of users, for the case of 4 RRHs and 128 subcarriers.

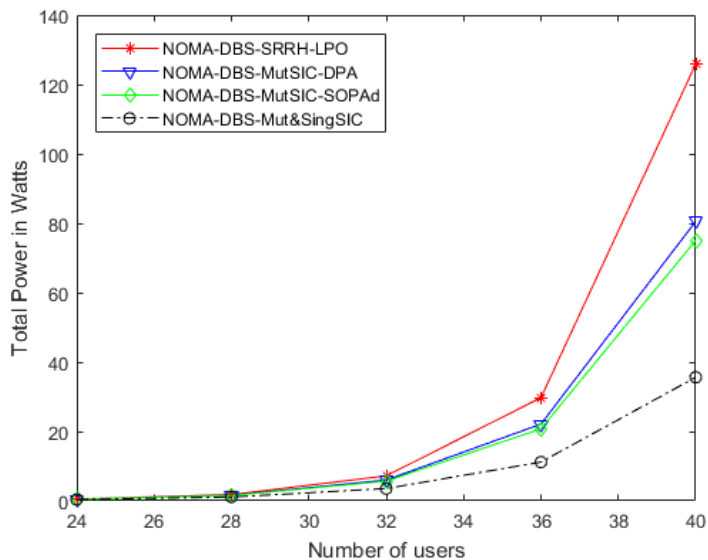

Fig. 5. Total power in terms of $K$, for $R_{k,req} = 5$ Mbps, $S = 128$ and $R = 4$.

Results prove that the allocation strategies based on mutual SIC, or combined mutual and single SIC, scale much better to crowded areas, compared to single SIC solutions. The power reduction of NOMA-DBS-Mut&SingSIC towards NOMA-DBS-SRRH-LPO is 62.4% and 71.7% for 36 and 40 users respectively.

In Table 2, we show the statistics of the number of non-multiplexed subcarriers, the number of subcarriers where a mutual SIC is performed, the number of subcarriers where a Single SIC is performed while the multiplexed users are powered by the same RRH, and the number of subcarriers where a Single SIC is performed while powering the paired users from different RRHs. On average, NOMA-DBS-SRRH-LPO uses Single SIC NOMA on 25% (resp. 32%) of the subcarriers for $R_{k,req} = 9$ Mbps (resp. 12 Mbps), while the rest of the subcarriers is mostly dedicated to a single user (a small proportion is not allocated at all, depending on the power threshold $\rho$). On the other hand, the proportions are respectively 7% and 9% with NOMA-DBS-MutSIC-SOPAd. Therefore, in the light of the results of figures 3 and 5, NOMA-DBS-MutSIC-SOPAd not only outperforms NOMA-DBS-SRRH-LPO from the power perspective, but it also presents the advantage of yielding a reduced complexity at the User Equipment (UE) level, by requiring a smaller amount of SIC procedures at the receiver side. This shows the efficiency of the mutual SIC strategy, combined with appropriate power adjustment, over classical single SIC configurations.

TABLE. 2. Statistics of user multiplexing, for $K = 15$, $S = 64$ and $R = 4$.

| Resource allocation technique | Non Mux SC | SC MutSIC | SC SingSIC SRRH | SC SingSIC DRRH |
|---|---|---|---|---|
| $R_{k,req} = 9$ Mbps | | | | |
| NOMA-DBS-SRRH-LPO | 48.156 | - | 15.844 | - |
| NOMA-DBS-MutSIC-SOPAd | 59.399 | 4.601 | - | - |
| NOMA-DBS-Mut&SingSIC | 39.332 | 4.601 | 4.984 | 15.083 |
| $R_{k,req} = 12$ Mbps | | | | |
| NOMA-DBS-SRRH-LPO | 43.203 | - | 20.797 | - |
| NOMA-DBS-MutSIC-SOPAd | 58.199 | 5.801 | - | - |
| NOMA-DBS-Mut&SingSIC | 28.64 | 5.801 | 7.489 | 22.07 |

It can be noted that in NOMA-DBS-Mut&SingSIC, 31% (resp. 44%) of the subcarriers are powered from different antennas, using either mutual or single SIC. This shows the importance of exploiting the additional spatial diversity, combined with NOMA, inherent to DBS systems.

## VII. CONCLUSION

In this paper, various resource allocation techniques were presented for minimizing the total downlink transmit power in DBS systems for 5G and beyond networks. We first proposed several enhancements to a previously developed method in the CBS context, prior to extending it to the DBS context. Furthermore, we unveiled the hidden potentials of DBS for NOMA systems and developed new techniques to make the most out of these advantages, while extracting their best characteristics and tradeoffs. Particularly, this study has enabled the design of NOMA with SIC decoding at both paired UE sides. Simulation results have shown the superiority of the proposed methods with respect to Single SIC configurations. They also promoted mutual SIC with suboptimal power adjustment to the best tradeoff between transmit power and complexity at both the PCC and the UE levels. Several aspects of this work can be further explored, since many additional challenges need to be addressed to enhance the NOMA-DBS-specific resource allocation schemes. For instance, the study can be enriched by the use of MIMO antenna systems, in a distributed context. Furthermore, practical considerations can be incorporated in the study, such as imperfect antenna synchronization and limited CSI exchange.

VIII. ACKNOWLEDGEMENT

This work has been funded with support from the Lebanese University and the franco-lebanese CEDRE program. Part of this work has been performed in the framework of the Horizon 2020 project FANTASTIC-5G (ICT-671660), which is partly funded by the European Union.